\documentclass[aps,pra,reprint,superscriptaddress]{revtex4-2}

\usepackage{amsmath,amsfonts,amssymb,mathrsfs,graphicx,xcolor}
\usepackage[colorlinks,linkcolor=blue,citecolor=blue,urlcolor=blue]{hyperref}

\renewcommand\gg{\gamma}
\newcommand\del{\delta}
\newcommand\ep{\varepsilon}

\newcommand\e{\eta}
\newcommand\q{\theta}

\renewcommand\l{\lambda}

\renewcommand\r{\rho}

\newcommand\f{\phi}

\renewcommand\j{\psi}

\newcommand\J{\Psi}

\newcommand{\non}{\nonumber}
\newcommand{\oo}{\mathscr{O}}

\def\ket#1{\lvert#1\rangle}

\def\bra#1{\langle #1 \rvert}

\def\amp#1#2{\langle #1 \lvert #2 \rangle}

\def\avg#1#2#3{\langle #1 \lvert #2 \lvert #3 \rangle}

\renewcommand\d{\partial}
\newcommand\ra{\rightarrow}

\newcommand{\lan}{\langle}
\newcommand{\ran}{\rangle}
\newcommand{\no}{\nonumber}

\begin{document}
	
	
	\title{Generalised Geometric Phase}
	
	
	\author{Vivek M. Vyas}
	\email[]{vivek.vyas@iiitvadodara.ac.in}
	\affiliation{Indian Institute of Information Technology Vadodara, Government Engineering College, Sector 28, Gandhinagar 382028, India}
	
	
	\date{\today}
	
	\begin{abstract}
		A generalised notion of geometric phase for pure states is found and its physical manifestations are shown. An appreciation of the fact that the interference phenomenon manifests in the average of an observable, allows us to define the argument of the matrix element of an observable as a generalised relative phase. This identification naturally paves the way for defining an operator generalisation of the geometric phase following Pancharatnam. The notion of natural connection finds an appropriate operator generalisation, and the generalised geometric phase is found to be the (an)holonomy of the generalised connection. Like the usual geometric phase, it is shown that the generalised geometric phase also manifests as a global phase acquired by a quantum state in course of time evolution. This phase is found to contribute to the shift in the energy spectrum due to perturbation, and to the forward scattering amplitude in the scattering problem.
	\end{abstract}
	
	
	\maketitle

	\section{Introduction}
	
	It is well known that the notion of geometric phase came to prominence from the celebrated work of Berry \cite{berry1984}. Berry showed that a quantum state acquired a non-trivial global phase when the system was evolved in an adiabatic cyclic manner to return to its initial configuration. Importantly it was shown that the phase acquired was of purely geometric origin, and was dependent upon the curvature of the underlying Hilbert space. The importance of this concept was immediately appreciated by the scientific community, and a flurry of activity followed soon after this landmark work \cite{shapere1989, anandan1997, anandan1992}. Soon the notion of geometric phase was generalised to include scenarios wherein the adiabaticity is absent, by Aharonov and Anandan\cite{anandan1987}. Eventually it was further generalised by Samuel and Bhandari to scenarios wherein the requirement of unitarity and cyclicity were also absent \cite{samuel1}. The mathematical framework behind the occurrence of the geometric phase also got slowly unfolded \cite{bsimon1983, page1987,mukunda1993}. The grounding of the notion of geometric phase got even further cemented by the experimental confirmation of this notion in classical optical systems \cite{chyba1988, chiao1986, chiao1988} and quantum systems like diatomic molecules and spin half particles\cite{mead1992,zwanziger1990berry, bitter1987manifestation}. The relevance and importance of geometric phase in classical systems also received significant attention \cite{agarwal1990, hannay1985angle, chaturvedi1987berry, littlejohn1988, anandan1988geometric, pati1998adiabatic,de2004notion}. This notion has also been studied in the context of particle physics \cite{zee1984,sonoda1986berry,stone1988berry, stone1986born,stone2015}. The importance of geometric phase in understanding condensed matter systems was noted immediately after Berry's work \cite{stone1986born,zee1984, thouless1983, bsimon1983, mathur1991} and forms the foundation of the current understanding of the physics of topological materials and quantum Hall effect, as also that of exotic objects like anyons \cite{thouless1982,thouless1985, bernevig2013, shapere1989,pz,roy2021}.  
	
	As noted by Berry \cite{berry1990,shapere1989}, the concept of geometric phase had been anticipated independently by several preceding workers. One of the elegant ways of understanding the occurrence of the geometric phase in its generality, is due to Pancharatnam \cite{pancharatnam1956,ramaseshan1986,berry1987}, which is solely based on the phenomenon of interference. This line of thought has turned out to be rewarding and has paved the way for a general setting for the geometric phase, wherein the evolution of the system need not be unitary, adiabatic or cyclic \cite{samuel1,samuel2}. This development eventually led to a purely kinematic understanding of the geometric phase \cite{mukunda1993,mukunda2003null}. 
	
	The essence of Pancharatnam's approach to arrive at the geometric phase lies in the concept of relative phase, defined as $\text{Arg}\amp{A}{B}$, which provides a scheme of comparison of any two non-orthogonal unit vectors $\ket{A}$ and $\ket{B}$. The usage of relative phase originates from the fact that the interference extrema as observed in the squared norm $\amp{A+B}{A+B}$ of the superposed vector $\ket{A + B}$ is determined by the relative phase $\text{Arg}\amp{A}{B}$. For a set of three non-orthogonal unit vectors, Pancharatnam's definition of the geometric phase is simply the sum of the corresponding relative phases.

	It is worth recollecting that apart from the squared norm $\amp{A+B}{A+B}$, the interference 
	phenomenon also manifests in the average $\avg{A+B}{\mathscr{O}}{A+B}$ of any observable $\mathscr{O}$. The interference extrema in such case is captured by the argument of the matrix element of the observable $\text{Arg}\avg{A}{\oo}{B}$. Here we propose that this phase can be understood as an operator generalised relative phase, and can be used as a general prescription to compare the vectors $\ket{A}$ and $\ket{B}$. Employing this generalised relative phase, following Pancharatnam's trail, we are naturally led to the construction of a generalised notion of geometric phase.
	We find that for a given set of $N$ states, this generalised geometric phase can be constructed for any observable of interest in the system at hand. The generalised geometric phase is found to go over to the usual geometric phase when the observable is taken as identity. We show that the concepts like natural connection and null phase curves, which are used to understand the usual geometric phase, also find an appropriate generalisation. We explicitly show that the generalised geometric phase is the (an)holonomy of the generalised natural connection. 
	
	One might wonder if this generalised geometric phase has any physical relevance, or it is just a theoretical construct. Interestingly we find that the generalised geometric phase can manifest in any general quantum system as a global phase acquired in course of time evolution, in scenarios wherein the usual geometric phase is undefined. We also find that within the framework of time independent perturbation theory, the change in the energy levels due to the perturbation has a contribution due to the generalised geometric phase. Remarkably in the context of scattering theory, we find that the generalised geometric phase manifests in the total scattering cross section as well as in the forward scattering amplitude. 
	
	The notion of geometric phase as understood using the kinematic framework is briefly reviewed in  Section (\ref{kinematic}). In Section (\ref{ggp}), the notions of generalised relative phase and generalised geometric phase are introduced and studied. In Section (\ref{physical}) we discuss the physical manifestations of the generalised geometric phase, and end the article with the summary of the obtained results.

	\section{Review of the notion of Geometric Phase \label{kinematic}}
	
	The concept of geometric phase can be understood elegantly from the viewpoint of Pancharatnam, whose origin lies in the interference phenomenon \cite{pancharatnam1956,ramaseshan1986,samuel1,samuel2}. Consider that we have some general physical system given to us, for example classical light beam or quantum point particle, which admits a Hilbert space $\mathscr{H}$. We assume that the system can be prepared in two different states depicted by $\ket {\j_{1}}$ and $\ket {\j_{2}}$ respectively. Now if the system is allowed to be in a linearly superposed state $\ket {\j} = \ket {\j_{1}} + \ket {\j_{2}} $, then it is well known that the interference phenomenon manifests in the squared norm of $\ket{\j}$ (identified with probability in quantum systems and light intensity in optical systems) :
	\begin{align} \label{inter1} 
		\amp \j \j =  \amp {\j_{1}} {\j_{1}} + \amp {\j_{2}} {\j_{2}} + 2 |\amp {\j_{1}} {\j_{2}}| \cos \left( \text{Arg} \amp {\j_{1}} {\j_{2}} \right).
	\end{align}
	The phase $\text{Arg} \amp {\j_{1}} {\j_{2}}$ is the \emph{relative phase}, which gives rise to the periodic extrema in $\amp \j \j$. The relative phase captures only the phase difference between the two states, which is evident from the fact that under a global phase transformation $\ket {\j_{i}} \ra e^{i \l_{i}}\ket {\j_{i}}$ (here $i = 1, 2$) for some real values of $\l_{i}$, the relative phase changes by amount $(\l_2 - \l_1)$. However it must be kept in mind that the interference phenomenon in $\amp{\j}{\j}$ is non-existent if the two states $\ket {\j_{1}}$ and $\ket {\j_{2}}$ are orthogonal. The concept of relative phase was used by Pancharatnam to provide a prescription for comparing any two vectors in the Hilbert space. Two vectors are said to be ``in-phase" in the sense of Pancharatnam if $\amp {\j_{1}} {\j_{2}}$ is real and positive, that is, if the relative phase $\text{Arg} \amp {\j_{1}} {\j_{2}}$ is zero \cite{pancharatnam1956,samuel1}. 
	
	Now suppose we consider three possible states $\ket {\j_{1}}$, $\ket {\j_{2}}$ and $\ket {\j_{3}}$ of the system, all belonging to $\mathscr{H}$. Given that $\ket {\j_{1}}$ is in-phase with $\ket {\j_{2}}$, and $\ket {\j_{2}}$ is in-phase with $\ket {\j_{3}}$, it is natural to wonder if $\ket {\j_{1}}$ and $\ket {\j_{3}}$ are also required to be in-phase. It is seen that the in-phase property is non-transitive, and the states $\ket {\j_{1}}$ and $\ket {\j_{3}}$ in general need not be in-phase with one another. The three states will be in-phase with one another if the sum of the relative phases $\text{Arg} \amp {\j_{1}} {\j_{2}} + \text{Arg} \amp {\j_{2}} {\j_{3}} + \text{Arg} \amp {\j_{3}} {\j_{1}}$ vanishes. It can be checked that while the individual relative phase can be altered by global phase changes $\ket {\j_{j}} \ra e^{i \l_{j}}\ket {\j_{j}}$ (where $j=1,2,3$)
	for some real numbers $\l_j$, the sum total is unaltered. This sum was used by Pancharatnam \cite{pancharatnam1956,samuel1,samuel2,mukunda1993}, who called it \emph{the excess phase}, as a measure of deviation of these states from the mutual in-phase configuration. It is generally expressed as:  
	\begin{align} \label{gp3}
		\gg_{123} = \text{Arg} \left( \frac{\amp {\j_{1}} {\j_{2}} \amp {\j_{2}} {\j_{3}} \amp {\j_{3}} {\j_{1}}}{\amp {\j_{1}} {\j_{1}} \amp {\j_{2}} {\j_{2}} \amp {\j_{3}} {\j_{3}}} \right).
	\end{align}
	The invariance of $\gg_{123}$ under three independent global phase transformations establishes the geometric nature of this phase, and hence it is called the \emph{geometric phase} \cite{berry1987,mukunda1993,samuel1}. The geometric nature become even more apparent by working with the density matrices $\rho_{j} = \frac{\ket{\j_{j}} \bra{\j_{j}}}{\amp{\j_j}{\j_j}}$ corresponding to the states $\ket{\j_j}$, so that (\ref{gp3}) now reads:
	\begin{align}
		\gg = \text{Arg} \: \text{Tr} \left( \r_1 \r_2 \r_3 \right).
	\end{align}
	
	\begin{figure}
		\begin{center}
			\includegraphics[scale=0.3]{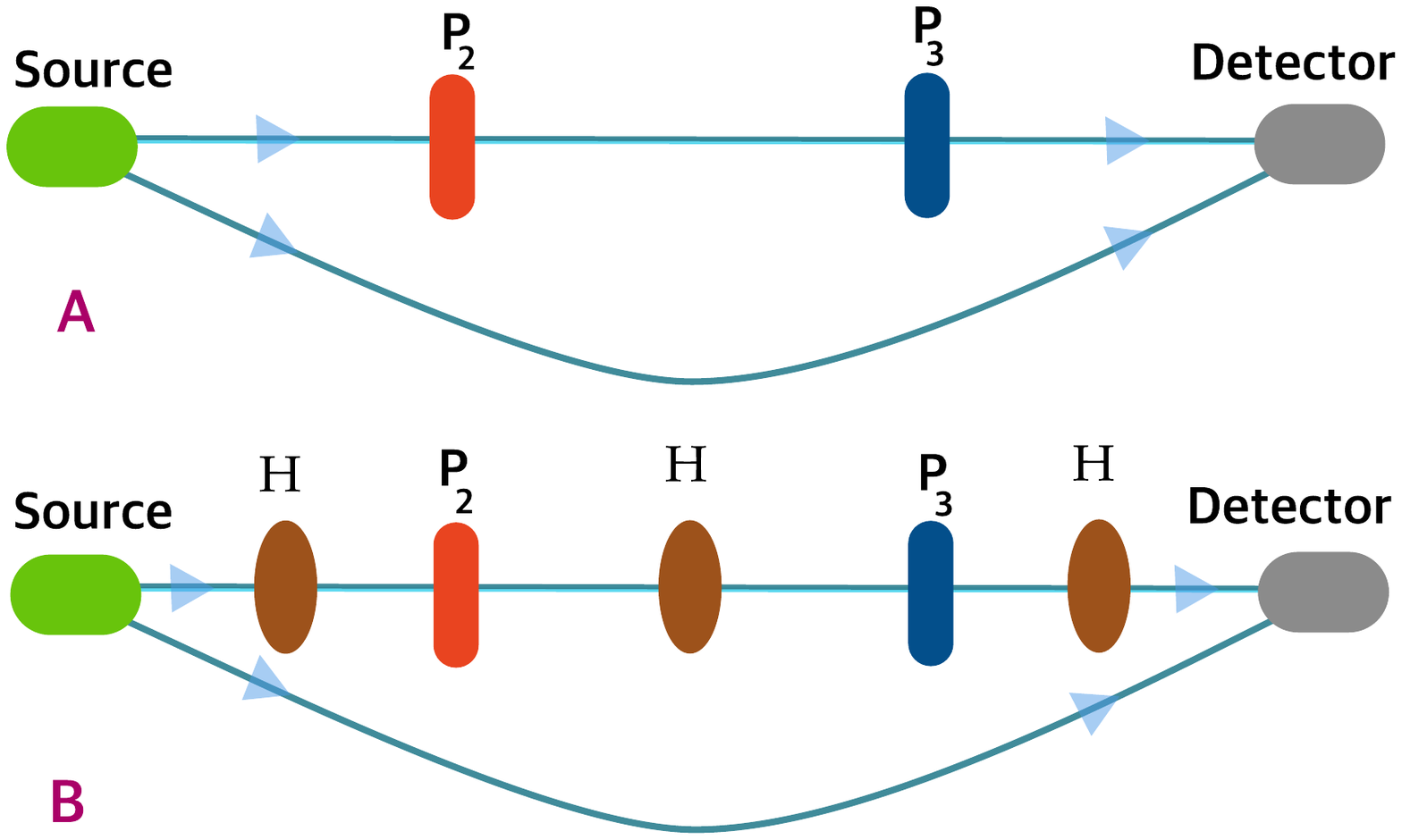}\caption{Schematic representation of the experimental setups to observe the geometric phases. Setup $A$ was used to measure the geometric phase $\gg$ whereas setup $B$ is designed to provide the measurement of generalised geometric phase $\gg_{H}$. Here $P_{2}$ and $P_{3}$ respectively are devices which project the system onto $\ket{\j_{2}}$ and $\ket{\j_{3}}$. The objects $H$ schematically represent the action of the medium on the system, which is akin to the generator of time evolution $H$. \label{fig1}}	
		\end{center}
	\end{figure}
	
	An experimental scheme based on interferometry to observe such a geometric phase was suggested in Ref. \cite{samuel1} and it is worth recollecting it here.
	As schematically depicted in setup $A$ in Fig. \ref{fig1}, suppose one prepares a quantum system at hand in state $\ket{\j_{1}}$ using some source. The setup is so arranged that there is negligible state evolution in the course of the experiment. Subsequently the system is subjected to a measurement using a device $P_{2}$ projecting it along $\ket{\j_{2}}$ followed by a measurement along $\ket{\j_{3}}$ using device $P_{3}$. In such a scenario the state of the system is $\ket{\j_{3}} \amp{\j_{3}}{\j_{2}} \amp{\j_{2}}{\j_{1}}$. Now if the system is allowed to interfere with the initial state $\ket{\j_{1}}$ then the relative phase measured by the detector turns out to be $\gg$. Interestingly such a geometric phase was experimentally observed using an optical interferometer in Ref. \cite{samuel2}, exploiting the polarisation states of light.

	This notion of geometric phase also holds for a collection of $N$ states $\{ \ket {\j_l} \}$ which belong to $\mathscr{H}$ (where $l = 1,2,\cdots , N$), and it reads:
	\begin{align} \label{gpN}
		\gg_N = \text{Arg} \left( \frac{\amp {\j_{1}} {\j_{2}} \amp {\j_{2}} {\j_{3}} \cdots \amp {\j_{N-1}} {\j_{N}} \amp {\j_{N}} {\j_{1}}}{\amp {\j_{1}} {\j_{1}} \amp {\j_{2}} {\j_{2}} \amp {\j_{3}} {\j_{3}}\cdots \amp {\j_{N}} {\j_{N}}} \right).
	\end{align}
	Evidently this notion of geometric phase can not be defined if any of the inner products in the numerator vanish when the states are orthogonal to each other.

	A special case of the above expression is when these $N$ states arise from the change of a continuous parameter $s$ ($0 \leq s \leq L$) such that $\ket {\j_{l}} \equiv \ket {\j(s_{l})}$, where $s_{l} = \frac{(l-1)}{(N-1)} \times L $. In large $N$ limit, the expression for the geometric phase reads:
	\begin{align}  \nonumber
		\gg(0,L) &= \text{Arg} \left( \frac{\amp {\j(L)} {\j(0)}}{\amp {\j(L)} {\j(L)}} \right) \\&+ \int_{0}^{L} \: ds \: \text{Im} \left( \frac{\amp {\j(s)} { d_{s}\j(s)}}{\amp {\j(s)} {\j(s)}}\right), \label{gpcont}
	\end{align}
	where $\ket {d_s \j(s)} \equiv \frac{d}{ds} \ket {\j(s)}$. The object $A_{\j}(s) = \text{Im} \left( \frac{\amp {\j(s)} { d_{s}\j(s)}}{\amp {\j(s)} {\j(s)}}\right)$ is often referred to as Berry potential \cite{shapere1989} and also as the natural connection in the literature \cite{bsimon1983,mukunda1993,samuel1}.  
	
	It can be readily checked that this geometric phase is invariant under local gauge transformation: $\ket {\j(s)} \ra e^{i \l(s)}\ket {\j(s)}$, for any function $\l(s)$, since the connection transforms as $A_{\j}(s) \rightarrow A_{\j}(s) + d_{s} \l(s)$. It can also be readily checked that the geometric phase is invariant under the reparametrization $s \ra r(s)$ (here $r(s)$ is a monotonically increasing function) such that $\ket {\j(s)} = \ket {\f(r)}$. These two invariances are a testimony to the geometric nature of this phase \cite{samuel1,mukunda1993}. It must be noted that this notion of geometric phase is \emph{not defined} if the initial and final states are orthogonal, as also if any two states $\ket {\j(s_l)}$ and $\ket {\j(s_{l+1})}$ are orthogonal. This is because our notion of relative phase fails for two linearly superposed orthogonal states.
	
	A very important and interesting scenario happens when the states $\ket{\j(s)}$ occur as an instantaneous eigenstate of the Hamiltonian $H(s)$, that is $H(s) \ket{\j(s)} = E(s) \ket{\j(s)}$, where $s$ is some slowly varying parameter in the system. 
	Further if the Hamiltonian is cyclic in $s$ so that $H(0) = H(L)$, then one wonders what happens to the system when it is initially prepared in $\ket{\j(0)}$, and $s$ is changed from $0$ to $L$ adiabatically. In a celebrated work Berry \cite{berry1984,berry1987} showed that the system will indeed return to $\ket{\j(0)}$ after a circuit, albeit with an overall phase which has a geometric component give by $\gg(0,L)$ as expressed in (\ref{gpcont}). It is evident that if the state $\ket{\j(s)}$ changes in such a manner that through out the circuit it is parallel to itself, which is ensured if $\text{Im} \amp {\j(s)} { d_{s}\j(s)} = 0$, then the relative phase $\text{Arg} \amp {\j(L)} {\j(0)}$ is the geometric phase $\gg(0,L)$ \cite{bsimon1983,anandan1992}. 
	
	It must be noted that the expression (\ref{gpcont}) is a general expression for the geometric phase acquired by the state of the system, and it is applicable to scenarios wherein the state evolution is neither cyclic, adiabatic or unitary \cite{anandan1987,samuel1,mukunda1993}. 
	
	Suppose one is given a set of states $\ket{\e(x)}$ where $0 \leq x \leq \tau$, which defines a continuous curve in the Hilbert space, such that the geometric phase $\gg(0,\tau) = 0$. We have in such a case:
	\begin{align} \label{nullgp}
		\text{Arg} \left( \frac{\amp {\e(0)} {\e(\tau)}}{\amp {\e(\tau)} {\e(\tau)}} \right)
		= \int_{0}^{\tau} dx \: \text{Im} \left( \frac{\amp {\e(x)} { d_{x} \e(x)}}{\amp {\e(x)} {\e(x)}}\right). 
	\end{align}
	Such curves are of special relevance in the understanding of geometric phase, and are called the \emph{null phase curves }\cite{mukunda2003null, chaturvedi2013null, rabei1999, goyal2022null}. The above expression tells us that the relative phase $\text{Arg} \amp{\e(0)}{\e(\tau)}$ is expressible as a line integral of the natural connection over the null phase curve. A classic result of Ref. \cite{samuel1} shows the existence and construction of such a curve connecting any two non-orthogonal states $\ket{A}$ and $\ket{B}$ in the Hilbert space \footnote{In Ref. \cite{samuel1} the curve connecting the two non-orthogonal states was found to be a geodesic corresponding to the Fubini-Study metric. It was later shown in Ref. \cite{mukunda2003null} that geodesic curve is a special case of null phase curve.}. It was shown that there exists a curve $\ket{\bar{\e}(x)}$ where:
	\begin{align}
		\ket{\bar{\e}(x)} = \frac{e^{i \theta x/\tau}}{\sin \tau} \left(  \sin (\tau - x) \ket{A} + e^{- i \theta} \sin (x) \ket{B} \right),
	\end{align}
	such that the relative phase between the two states $\theta = \text{Arg} \amp{B}{A}$ can be expressed as an integral of natural connection $\text{Im} \left( \frac{\amp {\bar{\e}(x)} { d_{x} \bar{\e}(x)}}{\amp {\bar{\e}(x)} {\bar{\e}(x)}}\right)$ along the curve $\ket{\bar{\e}(x)}$ connecting them \cite{samuel1, mukunda1993, mukunda2003null}.

	\begin{figure}
		\begin{center}
			\includegraphics[scale=0.5]{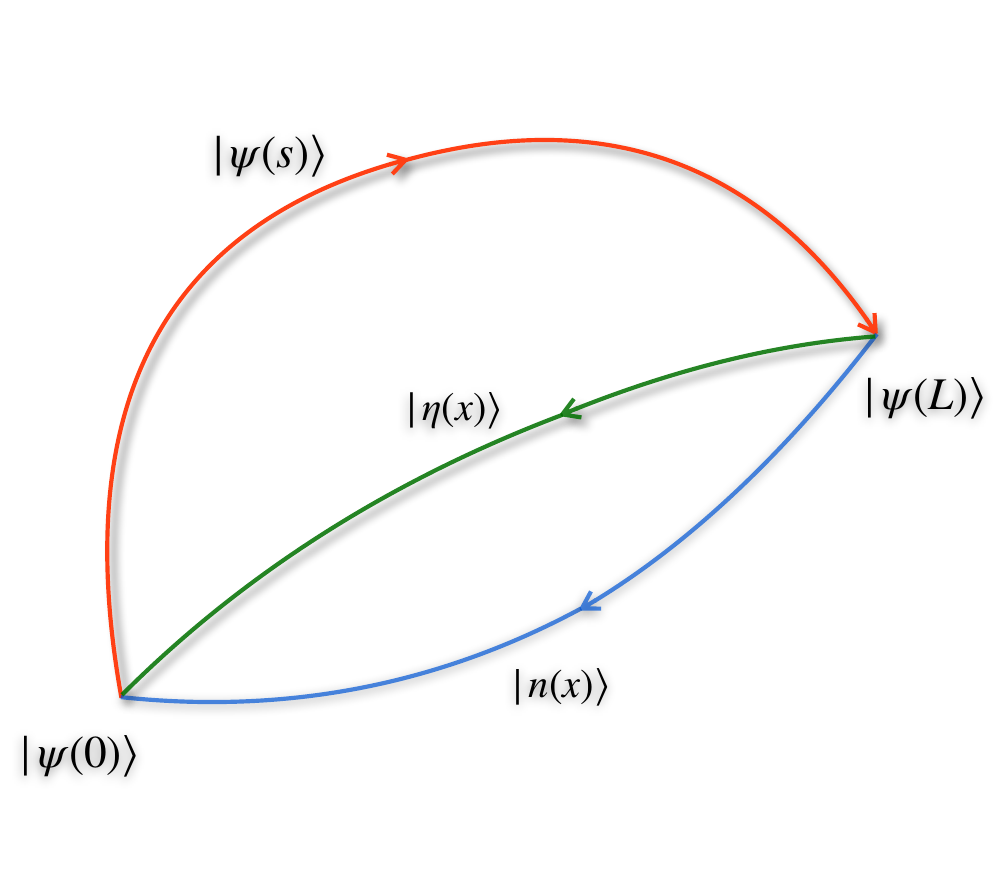}\caption{\label{intcircuit}Schematic representation of the integration path in the Hilbert space for evaluating the geometric phase. For evaluating $\gg(0,L)$ the path is along $\ket{\j(x)}$ (in red colour) and the null phase curve $\ket{\e(x)}$ (in green colour).  While for evaluating $\gg_{\oo}(0,L)$ the path is along $\ket{\j(x)}$ and $\oo$ null phase curve $\ket{n(x)}$ (in blue colour).}
		\end{center}	
	\end{figure}
	
	The special importance of these null phase curves also stems from the fact that if one is provided with a continuous curve $\ket{\j(s)}$ ($0 \leq s \leq L$) for which the geometric phase is $\gg(0,L)$, and a null phase curve $\ket{\e(x)}$ (where $0 \leq x \leq \tau$)  such that $\ket{\e(0)} \equiv \ket{\j(L)}$ and $\ket{\e(\tau)} \equiv \ket{\j(0)}$, then $\gg(0,L)$ can be expressed as a closed loop integral of the natural connection:
	\begin{align} \nonumber
		\gg(0,L) &= \int_{0}^{L} ds \: A_{\j}(s) + \int_{0}^{\tau} dx \: A_{\e}(x) \\
		& \equiv \oint_{C} dl \: A(l). 
	\end{align}
	Here the integrands are given by $A_{\j}(s) = \text{Im} \left( \frac{\amp {\j(s)} { d_{s}\j(s)}}{\amp {\j(s)} {\j(s)}}\right)$ and $A_{\e}(x) = \text{Im} \left( \frac{\amp {\e(x)} { d_{x} \e(x)}}{\amp {\e(x)} {\e(x)}}\right)$. The curve $C$ is specified from $\ket{\j(0)}$ to $\ket{\j(L)}$ through $\ket{\j(s)}$, whereas from $\ket{\j(L)}$ to $\ket{\j(0)}$ through $\ket{\e(x)}$, as depicted in Fig. \ref{intcircuit}. This expression clearly shows that the geometric phase is actually the (an)holonomy of the natural connection \cite{berry1990, anandan1992}. {It is also possible to understand the occurrence and the origin of geometric phase in the language of fibre bundles, as discussed in the Appendix.}      
	
	{The above closed integral can be written in an alternative form employing the Stokes theorem. Let $S$ be the surface formed by states $\ket{\j(k,j)}$ that is bounded by $C$ and is function of coordinates $k$ and $j$. In such a case it can be immediately seen that:
		\begin{align}
			\gg(0,L) &= \oint_{C} dl \: A(l) \\
			&= \int_{S} \: da \wedge db \: F_{ab} \equiv \int_{S} \: F  
		\end{align}
		where the curvature $F_{ab} = \partial_{a} A_{b} - \partial_{b} A_{a}$, (where $a,b$ take values $k,j$) and $A_{a} = \text{Im} \left( \frac{\amp {\j} { \partial_{a}\j}}{\amp {\j} {\j}}\right)$. Defining the covariant derivative as $D_{a}\ket{\j} = \partial_{a} \ket{\j} - i A_{a} \ket{\j} \equiv \ket{D_{a} \j}$, allows one to write the curvature as:
		\begin{align}
			F_{ab} = 2 \: \text{Im} \amp {D_{a} \j} {D_{b} \j}.
		\end{align}
		Often in the contemporary literature the quantity: 
		\begin{align}
			Q_{ab}&=\amp {D_{a} \j} {D_{b} \j} \\
			&=\frac{1}{\amp {\j}{\j}} \bra{\partial_a \j} \left( 1 - \ket{\j}\bra{\j}
			\right) \ket{\partial_b \j},
		\end{align}
		is referred to as the \emph{Quantum Geometric Tensor} \cite{gianfrate2020,provost1980}. It must be noted that $Q_{ab}$ by construction is gauge invariant, and hence is actually defined over the space of pure state density matrices $\rho_{\j} = \frac{\ket{\j} \bra{\j}}{\amp{\j}{\j}}$.
		
		The above observation raises a natural question so as to what is the significance of the real part of the geometric tensor. This was answered in a landmark work in 1980 by Provost and Vallee \cite{provost1980}, who showed that the real part of the geometric tensor provides one with a \emph{Riemannian metric} $g_{ab}$:
		\begin{align} \label{metricdef}
			g_{ab} &= \text{Re}\: Q_{ab},
		\end{align}
		albeit not over the Hilbert space, but over the space of pure state density matrices.
		Thus the infinitesimal distance squared $dL^2$ between two states $\ket{\j(s)}$ and $\ket{\j(s+ds)}$ on any curve parametrised by coordinate $s$ is given by:
		\begin{align}
			dL^2 &= g_{ss} ds^2.
		\end{align}  
		It is straightforward to see that this metric is the same as the celebrated \emph{Fubini-Study} metric \cite{nakahara2003,page1987,samuel1,bengtsson2017}, by observing that definition (\ref{metricdef}) can be arrived from the following expression when one considers two infinitesimally separated unit trace pure state density matrices $\r_{1} = \ket{\j(s)}\bra{\j(s)}$ and $\r_{2} = \ket{\j(s+ds)}\bra{\j(s+ds)}$:
		\begin{align} \label{fsmetric}
			dL^2 = 1 - \text{Tr}\: (\r_1 \r_2).
		\end{align}
		
		It is instructive to consider the case of a two level quantum system, whose Hilbert space consists of vectors $z_{1} \ket{0} + z_{2} \ket{1}$, where $\{z_1, z_2 \}$ are a pair of complex numbers. In this case it is easy to evaluate the quantum geometric tensor, and in terms of the ratio $w = z_2 / z_1$ one finds that the curvature $F$ and the metric respectively read \cite{nakahara2003,page1987,provost1980}:
		\begin{align}
			F &= - i \frac{dw \wedge d \bar{w}}{(1 + \bar{w} w)^2}, \\
			dL^2 &= \frac{dw d \bar{w}}{(1 + \bar{w} w)^2}.
		\end{align} 
		It the literature the above metric is known to represent a sphere in terms of complex coordinates \cite{nakahara2003}. This is not surprising since the space of pure state density matrices, also known as the ray space, in this example is known to be a complex projective space $CP^{1}$, which is isomorphic to a sphere \cite{nakahara2003,bengtsson2017}.}

	\section{Generalised Geometric Phase \label{ggp}}
	
	\subsection*{Generalisation of relative phase}
	
	In the earlier section, we began our discussion on the concept of interference by investigating the behaviour of $\amp {\j} {\j}$ as a function of relative phase $\text{Arg} \amp {\j_{1}} {\j_{2}}$ (as seen in (\ref{inter1})). It is on this edifice that most of the interference phenomena are understood in quantum physics as also in optics \cite{ sakurai1995modern,agarwal2012quantum, born2013, feynmanbook, ecgbook}. A careful reflection reveals that the phenomena of interference has manifestations beyond the probability or intensity. Apart from squared norm $\amp {\f}{\f}$, nature permits us to measure the average $\lan \mathscr{O} \ran_{\f} = \frac{\avg {\f} {\mathscr{O}} {\f}}{\amp {\f} {\f}}$ corresponding to an observable $\mathscr{O}$ when the system at hand is in the state $\ket \f$ \footnote{Here we have assumed that an observable $\mathscr{O}$ is a well defined self adjoint linear operator on the Hilbert space of the system at hand, while respecting the boundary conditions of the system.}. The average of observable $\mathscr{O}$ in a unit normalised superposed state $\ket{\j} = \ket{\j_{1}} +  \ket{\j_{2}}$, is given by:
	\begin{align} \non
		\lan \mathscr{O} \ran_{\j} &= {\avg{\j}{\mathscr{O}}{\j}}\\ \non
		& = \avg{\j_{1}}{\mathscr{O}}{\j_{1}} + \avg{\j_{2}}{\mathscr{O}}{\j_{2}}  \\  &+ 2 | \avg{\j_{1}}{\mathscr{O}}{\j_{2}} | \cos \left( \text{Arg} \avg{\j_{1}}{\mathscr{O}}{\j_{2}} \right). \label{intero}
	\end{align}
	Evidently the average of $\oo$ experiences periodic extrema as a function of phase $\text{Arg} \avg{\j_{1}}{\mathscr{O}}{\j_{2}}$. The phase $\text{Arg} \avg{\j_{1}}{\mathscr{O}}{\j_{2}}$ is sensitive to the relative phase change between the states $\ket{\j_{1}}$  and $\ket{\j_{2}}$, which can be seen from the fact that under a transformation $\ket{\j_{2}} \ra e^{i \q}\ket{\j_{2}}$ it changes by amount $\q$. As a result, the phase $\text{Arg} \avg{\j_{1}}{\mathscr{O}}{\j_{2}}$ can be thought of as a generalisation of relative phase $\text{Arg} \amp{\j_{1}}{\j_{2}}$. While the relative phase $\text{Arg} \amp{\j_{1}}{\j_{2}}$ governs the interference extrema in $\amp{\j}{\j}$, it is the relative phase $\text{Arg} \avg{\j_{1}}{\mathscr{O}}{\j_{2}}$ that determines the extrema in the average $\lan \mathscr{O} \ran_{\j}$. 
	
	Here we have crucially assumed that the two states $\ket{\j_{1}}$ and $\ket{\j_{2}}$ do not lie in the kernel of $\oo$, or else the interference phenomenon would be absent.
	
	It is clear that there is no definite relation between the two relative phases, since the amplitudes $\amp{\j_{1}}{\j_{2}}$ and $\avg{\j_{1}}{\mathscr{O}}{\j_{2}}$ are independent of one another. It is tempting to think of matrix element $\avg{\j_{1}}{\mathscr{O}}{\j_{2}}$ as the inner product $\amp{\j'_{1}}{\j'_{2}}$ between $\ket{\j'_1} = \sqrt{\oo} \ket{\j_1}$ and $\ket{\j'_2} = \sqrt{\oo} \ket{\j_2}$. One must however bear in mind that the square root $\sqrt{\oo}$ is in general ill defined, since it only makes sense for positive observables (whose all the eigenvalues are positive definite). In any case in general the overlap $\amp{\j_{1}}{\j_{2}}$ and the matrix element $\avg{\j_{1}}{\mathscr{O}}{\j_{2}}$ are fundamentally different physical and mathematical quantities.

	One can now use this \emph{$\oo$ relative phase} $\text{Arg} \avg{\j_{1}}{\mathscr{O}}{\j_{2}}$ as a criteria to define ``$\oo$ in-phase'' condition: the two states $\ket{\j_{1,2}}$ are said to be ``$\oo$ in-phase'' with one another if $\text{Arg} \avg{\j_{1}}{\mathscr{O}}{\j_{2}}$ is zero. It is evident that the two states can be $\oo$ in-phase but they need not be $\mathscr{O}'$ in-phase for some non-trivial operator $\oo \neq \mathscr{O}'$, as also the two states need not be in-phase in the Pancharatnam sense with $\amp{\j_{1}}{\j_{2}}$ being real and positive. It must be noted that the relative phase $\text{Arg} \amp{\j_{1}}{\j_{2}}$  is a special case of $\oo$ relative phase with $\oo = \mathbf{1}$.
	
	Interestingly in the literature the ratio of the two amplitudes $\amp{\j_{1}}{\j_{2}}$ and $\avg{\j_{1}}{\oo}{\j_{2}}$ has received significant attention and it is called the weak value \cite{dressel2014, aharonov1988, ecg1989} for observable $\mathscr{O}$:
	\begin{align}
		w_{\oo}(\j_{1},\j_{2}) = \frac{\avg{\j_{1}}{\oo}{\j_{2}}}{\amp{\j_{1}}{\j_{2}}}.	
	\end{align}
	In many experiments such weak values have been experimentally measured \cite{dressel2014, ritchie1991, hosten2008observation, bliokh2012}.   
	Evidently the weak value in general is a complex number, and hence its interpretation as an observable quantity within the framework of quantum mechanics has been intensely studied and debated \cite{jozsa2007, dressel2014, vaidman2017weak, hariri2019, berry2010typical, berry2011pointer}. In light of our conception of $\oo$ generalised relative phase 
	$\text{Arg} \avg{\j_{1}}{\mathscr{O}}{\j_{2}}$, one immediately sees another physical significance of the argument of the weak value $w_{\oo}(\j_{1},\j_{2})$, as the difference of the two relative phases:
	\begin{align}
		\text{Arg} \: w_{\oo}(\j_{1},\j_{2}) = \text{Arg} \avg{\j_{1}}{\mathscr{O}}{\j_{2}} 
		- \text{Arg} \amp{\j_{1}}{\j_{2}}.
	\end{align}
	If the argument of weak value is zero, then it tells us that the two relative phases are identical, and hence the interference manifestation in squared norm $\amp {\f}{\f}$ and the average $\lan \mathscr{O} \ran_{\j}$ is essentially identical. This indicates that the net impact of $\oo$ in defining the relative phase $\text{Arg} \avg{\j_{1}}{\mathscr{O}}{\j_{2}}$ is akin to identity. On the other hand, a non-zero value of the argument of weak value implies that the two relative phases are distinct, owing to the non-trivial role played by $\oo$. Thus the argument of weak value actually provides a measure of contribution of $\oo$ to the interference phenomenon as observed in $\lan \mathscr{O} \ran_{\j}$.  
	
	{It is worth mentioning that the connection between the usual geometric phase and the weak value of an observable has received significant interest \cite{sjoqvist2006} and it has been explored both theoretically and experimentally in several works, in particular, in the context of quantum erasers \cite{kobayashi2011,tamate2009,shikano2009,sjoqvist2006,ferraz2022}. Defining state $\ket{\j_{j'}}$ as:
		\begin{align*}
			\ket{\j_{j'}} = \frac{\oo \ket{\j_{j}}}{\sqrt{ \avg{\j_j}{\oo^2}{\j_j}}},
		\end{align*}
		enables us to write the weak value $\text{Arg} \: w_{\oo}(\j_{i},\j_{j})$ in terms of the geometric phase $\gg_{ij'j}$:
		\begin{align}\no
			\text{Arg} \: w_{\oo}(\j_{i},\j_{j}) &= \text{Arg} \amp{\j_i}{\j_{j'}}\amp{\j_{j'}}{\j_{j}}\amp{\j_j}{\j_{i}} \\ \no &- \text{Arg} \avg{\j_j}{\oo}{\j_j}\\
			&= \gg_{ij'j}  - \text{Arg} \avg{\j_j}{\oo}{\j_j}. \label{weakgeo}
		\end{align}
		
	}

	\subsection*{Generalisation of Geometric Phase}   
	
	With the above generalisation of the notion of relative phase, we are now in the position to generalise the concept of geometric phase, following the foot steps of Pancharatnam. Suppose we are given three possible unit normalised states $\ket {\j_{1}}$, $\ket {\j_{2}}$ and $\ket {\j_{3}}$ all belonging to $\mathscr{H}$. And we are also provided that $\ket {\j_{1}}$ is $\oo$ in-phase with $\ket {\j_{2}}$, and $\ket {\j_{2}}$ is $\oo$ in-phase with $\ket {\j_{3}}$. We now ask whether it implies that $\ket {\j_{1}}$ and $\ket {\j_{3}}$ are also $\oo$ in-phase ? It can be easily checked that the $\oo$ in-phase property is non-transitive, and the states $\ket {\j_{1}}$ and $\ket {\j_{3}}$ in general need not be $\oo$ in-phase with one another. The measure of deviation of these states from mutual $\oo$ in-phase configuration will now be captured by:
	\begin{align} \label{ggpthree}
		\gg^{123}_{\oo} &= \text{Arg} \left( \frac{ \avg{\j_{1}}{\mathscr{O}}{\j_{2}} \avg{\j_{2}}{\mathscr{O}}{\j_{3}} \avg{\j_{3}}{\mathscr{O}}{\j_{1}}}{{\avg{\j_{1}}{\mathscr{O}}{\j_{1}}} \avg{\j_{2}}{\mathscr{O}}{\j_{2}} \avg{\j_{3}}{\mathscr{O}}{\j_{3}}}\right) \\
		&= \text{Arg} \: \frac{\text{Tr} \left( \r_1 \oo \r_2 \oo \r_3 \oo \right)}{ \text{Tr} (\r_1 \oo) \text{Tr} (\r_2 \oo) \text{Tr} (\r_3 \oo)}.
	\end{align}
	This is the operator generalisation of the concept of geometric phase $\gg$ as defined in (\ref{gp3}). Evidently the operator generalised geometric phase $\gg^{123}_{\oo}$ goes over to the usual geometric phase $\gg_{123}$ when $\oo = \mathbf{1}$. It can be readily checked that the phase $\gg^{123}_{\oo}$ is invariant under the three independent global phase transformations: $\ket {\j_{j}} \ra e^{i \l_{j}}\ket {\j_{j}}$ (where $j=1,2,3$) for some real numbers $\l_j$, which is a proof of the geometric nature of $\gg_{\oo}$. 
	
	{
		It must be mentioned that the phase $\gg^{123}_{\oo}$ is well defined and in general has a distinct value for each given observable $\oo$. Employing weak values and (\ref{weakgeo}), one finds a relation between the geometric phase $\gg_{123} (\equiv \gg_{\mathbf{1}})$
		and the generalised geometric phase $\gg^{123}_{\oo}$ as:
		\begin{align} \no
			\gg^{123}_{\oo} &= \text{Arg} \left( w_{\oo}(\j_{1},\j_{2}) w_{\oo}(\j_{2},\j_{3}) w_{\oo}(\j_{3},\j_{1})\right) + \gg_{123} \\
			& = \gg_{12'2} + \gg_{23'3} + \gg_{31'1} + \gg_{123}. \label{gprel}
		\end{align}
		It must be kept in mind that this relation is not valid for all the states $\ket{\j_{j}}$, since we have tacitly assumed that not only the states $\ket{\j_{j}}$ are mutually non-orthogonal,  also that all the geometric phases $\gg_{12'2}$, $\gg_{23'3}$ and $\gg_{31'1}$ are well defined. In case when any of the states $\ket{\j_{j}}$ and $\ket{\j_{j'}}$ are mutually orthogonal, this relation does not hold. Thus the generalised geometric phase can not always be traded off for the sum of usual geometric phases, as defined above, and vice versa.  This establishes the existence of the generalised geometric phase as a geometrical quantity in its own right.}

	The notion of generalised geometric phase can also be extended to a collection of $N$ states $\{ \ket {\j_l} \}$ which belong to $\mathscr{H}$ (where $l = 1,2,\cdots , N$), and it reads:
	\begin{widetext}
		\begin{align} \label{ogpN}
			\gg_{\oo,N} = \text{Arg} \left( \frac{ \avg{\j_{1}}{\mathscr{O}}{\j_{2}} \avg{\j_{2}}{\mathscr{O}}{\j_{3}} \cdots \avg{\j_{N-1}}{\mathscr{O}}{\j_{N}} \avg{\j_{N}}{\mathscr{O}}{\j_{1}}}{  \avg{\j_{1}}{\mathscr{O}}{\j_{1}} \avg{\j_{2}}{\mathscr{O}}{\j_{2}} \cdots \avg{\j_{N}}{\mathscr{O}}{\j_{N}}} \right).
		\end{align}
	\end{widetext}
	{Evidently this notion of geometric phase can not be defined if any of the amplitudes in the numerator of (\ref{ogpN}) or (\ref{ggpthree}) vanish, as also if any of the averages in the denominator vanish.}  
	
	Let us consider a special case wherein these $N$ states arise from the change of a continuous parameter $s$ ($0 \leq s \leq L$) such that $\ket {\j_{l}} \equiv \ket {\j(s_{l})}$, where $s_{l} = \frac{(l-1)}{(N-1)} \times L $ so that we have:
	\begin{align} \nonumber
		&\frac{ \avg{\j_{1}}{\mathscr{O}}{\j_{2}} \avg{\j_{2}}{\mathscr{O}}{\j_{3}} \cdots \avg{\j_{N-1}}{\mathscr{O}}{\j_{N}} \avg{\j_{N}}{\mathscr{O}}{\j_{1}}}{  \avg{\j_{1}}{\mathscr{O}}{\j_{1}} \avg{\j_{2}}{\mathscr{O}}{\j_{2}} \cdots \avg{\j_{N}}{\mathscr{O}}{\j_{N}}} \\
		& =   \frac{\avg{\j_{N}}{\oo}{\j_{1}}}{\avg{\j_{N}}{\oo}{\j_{N}}} \: 
		\exp \left\{ \int_{0}^{L} \: ds \: \frac{\avg {\j(s)} {\oo} { d_{s}\j(s)}}{\avg {\j(s)}{\mathscr{O}}{\j(s)} }  \right\}. 	
	\end{align}
	Here we have assumed that none of the averages $\avg{\j_{l}}{\oo}{\j_{l}}$ are vanishing. In light of this expression, the generalised geometric phase $\gg_{\oo}(0,L) = \lim_{N \rightarrow \infty}\gg_{\oo,N}$ can now be written as:  
	\begin{align} \non
		\gg_{\oo}(0,L) &= \text{Arg} \left( \frac{\avg {\j(L)} {\oo} {\j(0)}}{\avg {\j(L)}{\oo}{\j(L)}} \right) \\& + \int_{0}^{L} \: ds \: \text{Im} \left( \frac{\avg {\j(s)} {\oo} { d_{s}\j(s)}}{\avg {\j(s)}{\mathscr{O}}{\j(s)} }\right). \label{ogpcont}
	\end{align}
	One immediately sees that the object $A^{\oo}_{\j}(s) = \text{Im} \left( \frac{\avg {\j(s)} {\oo} { d_{s}\j(s)}}{\avg {\j(s)}{\mathscr{O}}{\j(s)} }\right)$ is an operator generalisation of the natural connection or the Berry potential $A^{\mathbf{1}}_{\j}(s) \equiv A_{\j}(s)$, encountered earlier. It can be readily checked that $\gg_{\oo}(0,L)$ is invariant under the local gauge transformation: $\ket {\j(s)} \ra e^{i \l(s)}\ket {\j(s)}$, for any smooth function $\l(s)$, owing to the fact that the generalised connection transforms as $A^{\oo}_{\j}(s) \rightarrow A^{\oo}_{\j}(s) + d_{s} \l(s)$. One can also see that the generalised geometric phase is also invariant under the reparametrization $s \ra r(s)$ (where $r(s)$ is a monotonically increasing function) such that $\ket {\j(s)} = \ket {\f(r)}$. It is also evident that $\gg_{\oo}(0,L)$ goes over to $\gg(0,L)$ when $\oo = \mathbf{1}$.
	
	{It must be emphasised that the generalised connection $A^{\oo}_{\j}(s)$ and the generalised geometric phase $\gg_{\oo}(0,L)$ are not well defined if the curve $\ket {\j(s)}$ traverses through a region in $\mathscr{H}$ wherein the average of $\oo$ vanishes $\avg{\j(s)}{\oo}{\j(s)} = 0$.}

	\subsection*{Null phase Curves}
	
	The notion of null phase curve can also be extended naturally to the generalised geometric phase $\gg_{\oo}$. For a given continuous curve $\ket{n(x)}$ (where $0 \leq x \leq \tau$) if the geometric phase 
	\begin{align} \nonumber
		\gg_{\oo}(0,\tau) &=  \text{Arg} \left( \frac{\avg {n(\tau)} {\oo} {n(0)}}{\avg {n(\tau)}{\oo} {n(\tau)}} \right) \\ & + \int_{0}^{\tau} \: dx \: \text{Im} \left( \frac{\avg {n(x)} {\oo} { d_{x} n(x)}}{\avg {n(x)}{\mathscr{O}}{n(x)} }\right),
	\end{align}
	is \emph{zero}, then we say that it is $\oo$ \emph{null phase curve}. Since the value of the geometric phase $\gg(0,\tau)$ (as given by (\ref{nullgp})) would in general be different from $\gg_{\oo}(0,\tau)$, a null phase curve corresponding to $\gg(0,\tau)$ need not be a $\oo$ null phase curve, and vice versa. 
	
	We now show from the above definition of the $\oo$ null phase curve, that for any given two states $\ket{A}$ and $\ket{B}$ in $\mathscr{H}$, there always exists a $\oo$ null phase curve that connects them, so long as the amplitude $\avg{A}{\oo}{B} \neq 0$. Without loss of generality, let us assume that 
	there exists a curve $\ket{n(x)}$ (where $0 \leq x \leq \tau$), such that $\ket{A} \equiv \ket{n(0)}$ and $\ket{B} \equiv \ket{n(\tau)}$. Let us perform a gauge transformation to define a new curve $\ket{N(x)} = e^{i \theta \frac{x}{\tau}} \ket{n(x)}$, where  $\theta = \text{Arg} \left( \frac{\avg{B}{\oo}{A}}{\avg{B}{\oo}{B}}\right)$. One sees that $\text{Arg} \left( \frac{\avg {N(\tau)} {\oo} {N(0)}}{\avg {N(\tau)}{\oo}{N(\tau)}}\right) = 0$, implying that $\avg {N(\tau)} {\oo} {N(0)}$ is real. Now let us define the curve as $\ket{N(x)} = (1 - \frac{x}{\tau}) \ket{N(0)} + \frac{x}{\tau} \ket{N(\tau)}$. One can straight away see that $\text{Im} \left( \frac{\avg {N(x)} {\oo} { d_{x} N(x)}}{\avg {N(x)}{\mathscr{O}}{N(x)} }\right) = 0$ for this curve. This shows that $\ket{N(x)}$ is a $\oo$ null phase curve albeit connecting $\ket{N(0)}= \ket{A}$ to $\ket{N(\tau)}= e^{i \theta}\ket{B}$. Undoing the gauge transformation one obtains the expression for the $\ket{n(x)}$ curve as:
	\begin{align}
		\ket{n(x)} = e^{- i \theta \frac{x}{\tau}} \left( (1 - \frac{x}{\tau}) \ket{A} + \frac{x}{\tau} e^{i \theta} \ket{B} \right).	
	\end{align}
	Since the property of having zero geometric phase is invariant under local gauge transformation, one infers that the curve $\ket{n(x)}$ is indeed a $\oo$ null phase curve, which connects $\ket{A}$ to $\ket{B}$. As a result we have that:
	\begin{align}
		\text{Arg} \left( \frac{\avg {A} {\oo} {B}}{\avg {B}{\oo}{B}} \right) = \int_{0}^{\tau} \: dx \: \text{Im} \left( \frac{\avg {n(x)} {\oo} { d_{x} n(x)}}{\avg {n(x)}{\mathscr{O}}{n(x)} }\right)	
	\end{align}

	\begin{figure}
		\begin{center}
			\includegraphics[scale=0.5]{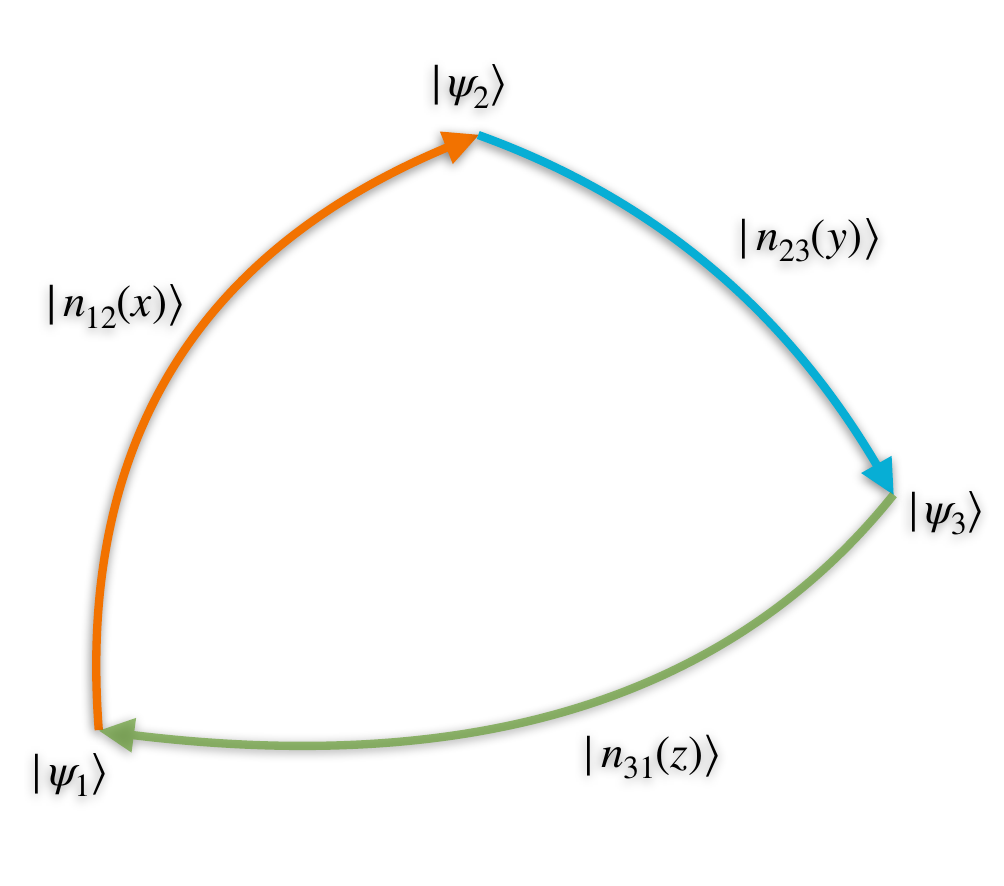}\caption{\label{threecircuit}Schematic representation of the integration path in the Hilbert space for evaluating $\gg_{\oo}$ as expressed in (\ref{ggpthree}).}	
		\end{center}
	\end{figure}
	
	The existence of the $\oo$ null phase curve provides one with a pleasing way to express $\gg_{\oo}(0,L)$ as a closed loop integral of the generalised natural connection. Consider that one is given a non-trivial curve $\ket{\j(s)}$ ($0 \leq s \leq L$) as depicted in Fig. \ref{intcircuit} connecting $\ket{\j(0)}$ to $\ket{\j(L)}$, for the geometric phase is $\gg_{\oo}(0,L)$. Invoking the above construction we obtain a $\oo$ null phase curve $\ket{n(x)}$ connecting $\ket{\j(L)} \equiv \ket{n(0)}$ to $\ket{\j(0)} \equiv \ket{n(\tau)}$, so that the geometric phase is expressible as:
	\begin{align} \nonumber
		\gg_{\oo}(0,L) = & \int_{0}^{L} \: ds \: {A}^{\oo}_{\j}(s) + \int_{0}^{\tau} \: dx \: {A}^{\oo}_{n}(x)  \\ & \equiv \oint_{\mathscr{C}} \: dl \: {A}^{\oo}(l).
	\end{align}
	Here we have defined the generalised connection as ${A}^{\oo}_{\j}(s) = \text{Im} \left( \frac{\avg {\j(s)} {\oo} { d_{s}\j(s)}}{\avg {\j(s)}{\mathscr{O}}{\j(s)} } \right)$, and ${A}^{\oo}_{n}(x) = \text{Im} \left( \frac{\avg {n(x)} {\oo} { d_{x} n(x)}}{\avg {n(x)}{\mathscr{O}}{n(x)} } \right)$. The closed curve $\mathscr{C}$ consists of $\ket{\j(s)}$ and $\ket{n(x)}$ as depicted in Fig. \ref{intcircuit}.
	
	Interestingly the existence of a $\oo$ null phase between any two states allows us to re-express the generalised geometric phase ${\gg_{\oo}}$ (as defined in (\ref{ggpthree})) as a closed loop integral along $\Delta$ formed by the three $\oo$ null phase curves $\ket{n_{12}(x)}$, $\ket{n_{23}(y)}$ and $\ket{n_{31}(z)}$ (see Fig. \ref{threecircuit}):
	\begin{align}
		\gg_{\oo} &= \int_{n_{12}} \: dx \: {A}^{\oo}_{12}(x) + \int_{n_{23}} \: dy \: {A}^{\oo}_{23}(y) + \int_{n_{31}} \: dz \: {A}^{\oo}_{31}(z) \\
		& \equiv \oint_{\Delta} \: dl \: {A}^{\oo}(l). 
	\end{align}
	Here the convention describing the $\oo$ null phase curve $\ket{n_{ij}}$ is that it connects $\ket{\j_{i}}$ to $\ket{\j_{j}}$, along which the connection is ${A}^{\oo}_{ij} = \text{Im} \left( \frac{\avg {n_{ij}(x)} {\oo} { d_{x} n_{ij}(x)}}{\avg {n_{ij}(x)}{\mathscr{O}}{n_{ij}(x)} } \right)$.
	
	\emph{This clearly shows that the geometric phase $\gg_{\oo}$ is the (an)holonomy of the generalised connection $A^{\oo}$}.

\begin{figure}
		\begin{center}
			\includegraphics[scale=0.5]{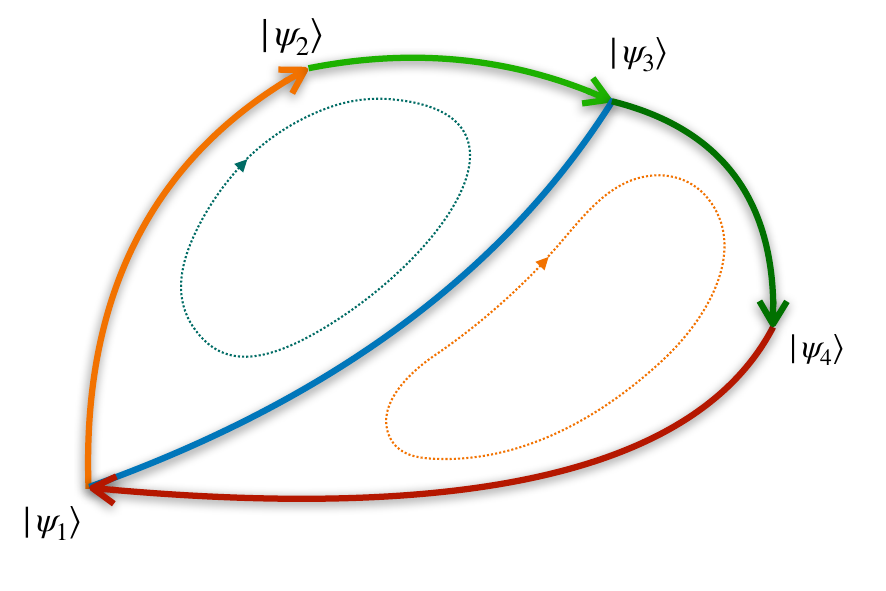}\caption{\label{gpadd} Schematic representation for the integration path for evaluating the generalised geometric phase for a set of four states $\ket{\j_{j}}$ where $j=1,2,3,4$. Here the solid curves connecting the states represent the respective null phase curves. The dotted curves represent the integration contour to evaluate the geometric phase for the triad of states.}	
		\end{center}
	\end{figure}

	The representation of the geometric phase as a closed loop integral also brings to fore its additive property akin to area. Suppose we are interested in calculating the generalised geometric phase $\gg_{\oo}(1,2,3,4)$ for a set of four states $\ket{\j_{j}}$ for $j=1,2,3,4$. The integration path $C$ is now along the null phase curves $\ket{n_{12}}$, $\ket{n_{23}}$, $\ket{n_{34}}$ and $\ket{n_{41}}$, as depicted in Fig. \ref{gpadd}. However by adding and subtracting the integral along null phase curve $\ket{n_{31}}$ it is evident that the integral can also be expressed as:   
	\begin{align}
		\gg_{\oo}(1,2,3,4) &= \oint_{C}  dl \: A^{\oo}(l) \\
		&= \oint_{\Delta_{123}}  dl \: A^{\oo}(l) + \oint_{\Delta_{134}}  dl \: A^{\oo}(l) \\
		&= \gg_{\oo}(1,2,3) + \gg_{\oo}(1,3,4).
	\end{align}
	Here the geometric phase $\gg_{\oo}(1,2,3) = \oint_{\Delta_{123}}  dl \: A^{\oo}(l)$ is obtained by evaluating the integral along curve $\Delta_{123}$ (represented as dotted green curve in Fig. \ref{gpadd}) which is formed by connecting the curves $\ket{n_{12}}$, $\ket{n_{23}}$ and $\ket{n_{31}}$, while $\gg_{\oo}(1,3,4) = \oint_{\Delta_{134}}  dl \: A^{\oo}(l)$ is obtained by evaluating the integral along curve $\Delta_{134}$ (represented as dotted orange curve in Fig. \ref{gpadd}) which is formed by connecting the curves $\ket{n_{13}}$, $\ket{n_{34}}$ and $\ket{n_{41}}$.

	One can extend this logic further to show that the generalised geometric phase for a collection of $N$ states $\{ \ket {\j_l} \}$ (where $l = 1,2,\cdots , N$) as given by (\ref{gpN}), can also be broken down as a sum of the generalised geometric phases due to the collection of triads $\{ \ket{\j_{1}}, \ket{\j_{i}}, \ket{\j_{i+1} }\}$ (for $2 \leq i \leq N-1$) as:
	\begin{align} 
		\gg_{\oo,N} = \sum_{i=2}^{N-1} \gg_{\oo}(1,i,i+1).
	\end{align}

	{By employing the Stokes theorem one can express the generalised geometric phase as a surface integral of the generalised curvature. For a given closed curve $\mathcal{C}$ defined by states $\ket{\j(l)}$, this can be expressed as:
		\begin{align}
			\oint_{\mathcal{C}} \: dl \: {A}^{\oo}(l) = \int_{S} da \wedge db \: F^{\oo}_{ab} \equiv \int_{S} \: F^{\oo}, 
		\end{align}
		where the curvature $F^{\oo}_{ab} = \d_{a} A^{\oo}_{b} - \d_{b} A^{\oo}_{a}$ is defined using the connection components ${A}^{\oo}_{s(t)} = \text{Im} \left( \frac{\avg {\j} {\oo} { d_{s(t)} \j } }{\avg {\j}{\mathscr{O}}{\j} } \right)$. Here the surface $S$ is bounded by $\mathcal{C}$, with parameters $s$ and $t$ specifying it. Defining the covariant derivative as $D_{a}\ket{\j} = \partial_{a} \ket{\j} - i A^{\oo}_{a} \ket{\j} \equiv \ket{D_{a} \j}$, allows one to write the curvature as:
		\begin{align}
			F^{\oo}_{ab} = 2 \: \text{Im} \avg {D_{a} \j}{\oo}{D_{b} \j}.
		\end{align}
		One can use this knowledge to define a \emph{Generalised Quantum Geometric Tensor} as: 
		\begin{align}
			Q^{\oo}_{ab}&=\avg {D_{a} \j}{\oo}{D_{b} \j} \\
			&=\frac{1}{\avg {\j}{\oo}{\j}} \bra{\partial_a \j} \left( 1 - \oo\ket{\j}\bra{\j} \oo
			\right) \ket{\partial_b \j}.
		\end{align}
		It must be noted that $Q_{ab}$ by construction is gauge invariant, and hence is actually defined over the space of pure state density matrices $\rho_{\j}$.
		
		It is straightforward to see that the real part of the generalised geometric tensor provides one with a \emph{pseudo-Riemannian metric} $g_{ab}$:
		\begin{align} \label{gmetricdef}
			g^{\oo}_{ab} &= \text{Re}\: Q^{\oo}_{ab},
		\end{align}
		over the space of pure state density matrices. We thus have a notion of infinitesimal distance squared $dl^2$ between two states $\ket{\j(s)}$ and $\ket{\j(s+ds)}$ on any curve parametrised by coordinate $s$ as:
		\begin{align}
			dl^2 &= g^{\oo}_{ss} ds^2.
		\end{align}  
		The definition (\ref{gmetricdef}) can be arrived from the following expression when one considers two infinitesimally separated pure state density matrices $\r_{1} = \ket{\j(s)}\bra{\j(s)}$ and $\r_{2} = \ket{\j(s+ds)}\bra{\j(s+ds)}$:
		\begin{align}
			dl^2 = 1 - \frac{\text{Tr}\: (\r_1 \oo \r_2 \oo)}{\text{Tr}\: (\r_1 \oo )\text{Tr}\: (\r_2 \oo)}.
		\end{align}
		This clearly shows that this indefinite metric $dl^2$ is an operator generalisation of Fubini-Study metric encountered earlier in (\ref{fsmetric}). 
		
		It will be useful to consider the case of a two level quantum system, whose Hilbert space consists of vectors $z_{1}\ket{0} + z_{2}\ket{1}$, with the observable $\oo = \sigma_{z} = \ket{0} \bra{0} - \ket{1}\bra{1}$. In this case the generalised quantum geometric tensor can be readily evaluated. In terms of the ratio $w = z_2 / z_1$ one finds that the generalised curvature $F^{\sigma_{z}}$ and the metric respectively read \cite{nakahara2003,page1987,provost1980}:
		\begin{align}
			F^{\sigma_{z}} &= - 2 i \frac{dw \wedge d \bar{w}}{(1 - \bar{w} w)^2}, \\
			dl^2 &= \frac{-2 dw d \bar{w}}{(1 - \bar{w} w)^2}.
		\end{align} 
		It the literature the above metric is known to represent a non-compact hyperboloid surface in terms of complex coordinates \cite{nakahara2003}. This brings to fore the impact of the presence of the observable $\sigma_{z}$ on the underlying geometry. It is clearly evident that the geometry is generally distinct when different observables are considered. For example, in the earlier section, we found that for quantum two level system, the geometry came out to be that of a sphere with Fubini-Study metric when observable was taken to be trivial $\oo = \mathbf{1}$. This shows the richness that a quantum system can display through the generalised geometric phase $\gg_{\oo}$ with a suitable choice of the observable \cite{vyas2023}.}   
	
	{Akin to the usual geometric phase case, the genesis and occurrence of the generalised geometric phase can also be clearly understood in the language of fibre bundles, as discussed in the Appendix.}
	
	\section{Physical manifestions of Generalised Geometric Phase \label{physical}}
	
	The reader might wonder whether the geometric phase $\gg_{\oo}$ is just another mathematical construct or if it provides some physical insight about the system at hand. After a careful study we find that the operator generalised geometric phase provides an alternative means of defining the coherence property of the set of states.

	\subsection*{Phase acquired by a state due to measurements}
	
	It is natural to ask if the generalised geometric phase can be understood as a phase acquired by a quantum state in course of evolution. To answer this let us assume that we are provided with a system defined by the time independent positive operator Hamiltonian $H = H_{0} + H_{\text{int}}$, where $H_{0}$ represents the free part of the Hamiltonian,  whereas $H_{\text{int}}$ stands for non-trivial interacting part of $H$. Let three unit normalised orthogonal eigenstates of $H_{0}$ be given by $\ket{0}$, $\ket{1}$ and $\ket{2}$. 
	
	Initially let the system be prepared in state $\ket{0}$ and allowed to evolve for some small time $\ep$. Subsequently let a projective measurement be performed on the system to find that it is in the state $\ket{1}$ (modulo an overall phase). The system is again allowed to evolve for time $\ep$ before getting subjected to a projective measurement which discovers that it is now in the state $\ket{2}$ (modulo an overall phase). Now after evolution for time $\ep$ if yet another measurement is performed so as to project it back to $\ket{0}$, then one finds that the state of the system is given by $\ket{0} \avg{0}{U(3 \ep, 2 \ep)}{2} \avg{2}{U(2 \ep, \ep)}{1} \avg{1}{U(\ep, 0)}{0}$. Here we are working in the Schrodinger picture of time evolution, wherein the evolution operator is $U(t_2,t_1) = \exp (- \frac{i}{\hbar} (t_2 - t_1) H)$. This shows that the system returns to its initial state albeit, to the leading order in $\ep$, acquiring a non-trivial operator generalised geometric phase:
	\begin{align}
		\gg_{H} = \text{Arg} \left( \avg{0}{H}{1} \avg{1}{H}{2} \avg{2}{H}{0}\right),
	\end{align}
	where the role of $\oo$ is played by $H$. This generalised geometric phase can also be experimentally measured by interferometry wherein the  system after time $3 \ep$, which is in the state $U(3\ep, 2\ep) \ket{2} \avg{2}{U(2 \ep, \ep)}{1} \avg{1}{U(\ep, 0)}{0}$, is allowed to be interfered to $\ket{0}$. This can be achieved by altering the interferometric setup $A$ depicted in Fig. \ref{fig1} to setup $B$ by ensuring that the states $|\j_{i}\rangle$ are chosen to be orthogonal (and are not eigenstates of $H$), and by introducing a suitable medium which acts on the states as the Hamiltonian operator $H$. As a result the relative phase measured by the detector would be $\gg_{H}$. It must be mentioned that  the usual geometric phase (as given by (\ref{ggpthree})) is not defined in this scenario since we are dealing with orthogonal states. 
	
	\emph{This clearly shows that operator generalised geometric phases $\gg_{\oo}$ are not mere theoretical constructs, but are as physical as the (identity) geometric phase $\gg$}.

	From the above discussion it is clear that the while there are many possible geometric phases $\gg_{\oo}$ for each choice of observable $\oo$, the physics of the given system determines so as to which geometric phase will be acquired by the state. While in the scenario considered above it is the phase $\gg_{H}$ that manifests, in the scenarios considered by Berry and others, it is the phase $\gg_{\mathbf{1}}$ that is physically relevant and experimentally observed.   
	
	\subsection*{Phase acquired by a state due to adiabatic evolution}
	
	{Let us consider a model non-Hermitian two level system, whose Hamiltonian is given by $H(\theta) = w_{1} e^{i \theta}\ket{0} \bra{1} + w_{2} e^{-i \theta}\ket{1} \bra{0}$, where $\theta$ is a real parameter. The energy eigenvalues are $E_{\pm} = \pm \sqrt{w_1 w_2}$, where both $w_1$ and $w_2$ are positive non-zero reals. It is well known that such a non-Hermitian Hamiltonian admits left and right eigenstates $\bra{\j_{\pm}^{L}(\theta)} H(\theta) = E_{\pm} \bra{\j_{\pm}^{L}(\theta)}$ and $H(\theta) \ket{\j_{\pm}^{R}(\theta)} = E_{\pm} \ket{\j_{\pm}^{R}(\theta)}$ which are given by $\bra{\j_{\pm}^{L}(\theta)} = \frac{1}{\sqrt{2}} ( \bra{0} \pm \sqrt{\frac{w_1}{w_2}} e^{i \theta}\bra{1})$ and $\ket{\j_{\pm}^{R}(\theta)} = \frac{1}{\sqrt{2}} ( \ket{0} \pm \sqrt{\frac{w_2}{w_1}} e^{-i \theta}\ket{1})$. Evidently the right and left eigenstates alone do not form an orthonormal basis set, but together they obey the bi-orthonormality relations: $\amp{\j_{\pm}^{L}(\theta)}{\j_{\pm}^{R}(\theta)}=1$ and $\amp{\j_{\pm}^{L}(\theta)}{\j_{\mp}^{R}(\theta)}=0$ \cite{garrison1988,roy2021,lieu2018topological}.
		
		Suppose if initially one prepares the system in the state $\ket{\j^{R}_{-}(0)}$ when $\theta = 0$. Following the footsteps of Berry, one can ask what happens to the system when the value of $\theta$ is adiabatically changed in a cyclic manner to $\theta = 2 \pi$. It was shown elegantly by Garrison and Wright \cite{garrison1988} that the system returns to its initial state picking up a geometric phase:  
		\begin{align}
			\gg_{-} = \int_{0}^{2 \pi} d\theta \: \mathrm{Im} \avg{\j_{-}^{L}(\theta)}{\frac{d}{d \theta}}{\j_{-}^{R}(\theta)},
		\end{align}
		which is a bi-orthonormal generalisation of Berry's result. This geometric phase has been found to successfully capture the topological properties of driven and dissipative condensed matter systems \cite{lieu2018topological,roy2021}, that are not captured by the usual geometric phase.
		
		Mostafazadeh \cite{mostafazadeh2010pseudo} has shown that such non-Hermitian system can be better dealt with by working with an inner product defined as: $\amp{A}{B}_{\eta}=\avg{A}{\eta}{B}$, where $\ket{A}$ and $\ket{B}$ are any two vectors. The operator $\eta$ is given by $\eta = \sum_{m} \ket{\j^{L}_{m}(t)} \bra{\j^{L}_{m}(t)}$, so that the states $\ket{\j_{n}^{R}(t)}$ now obey the $\eta$-orthonormality condition, which is also the bi-orthonormality condition: $\amp{\j_{m}^{R}(t)}{\j_{n}^{R}(t)}_{\eta} = \avg{\j_{m}^{R}(t)}{\eta}{\j_{n}^{R}(t)} = \amp{\j_{m}^{L}(t)}{\j_{n}^{R}(t)} =\del_{mn}$. In this light the above expression for the geometric phase interestingly becomes:
		\begin{align}
			\gg_{-} = \int_{0}^{2 \pi} d\theta \: \mathrm{Im} \avg{\j_{-}^{R}(\theta)}{\eta \frac{d}{d \theta}}{\j_{-}^{R}(\theta)}.
		\end{align}
		Here the operator $\eta = \ket{0}\bra{0} + \frac{w_1}{w_2} \ket{1}\bra{1}$. Thus we see that the geometric phase acquired by such a dissipative system is not the usual geometric phase given by Berry, rather it is the operator generalised geometric phase $\gg_{-} \equiv \gg_{\eta}$.
		
		Upon explicit evaluation it turns out that $\gg_{-} = -\pi$ if the system traverses the complete circuit from $\theta=0$ to $\theta=2\pi$. If it does not traverse a complete circuit and returns to $\theta=0$ from lets say $\theta=\pi$, then the geometric phase vanishes $\gg_{-} = 0$. This clearly shows that this geometric phase essentially is quantised, it only depends on the path topology and not on the finer details of the system, like parameters $w_{1,2}$. 
		
		Thus one sees that the operator generalised geometric phase is the geometric phase that is acquired by the state of such a dissipative system when it is made to adiabatically and cyclically return.}

	\subsection*{Effect on Energy Levels}

	Apart from the above manifestation, the generalised geometric phase also contributes to the spectrum of the quantum system due a time independent perturbation. Consider that one is given some general quantum system defined by Hamiltonian $H_{0}$, with eigenstates $\ket{n^{0}}$: $ H_{0} \ket{n^{0}} = E_{n}^{0} \ket{n^{0}}$. The system is subjected to a time independent perturbation so that the Hamiltonian of the system now reads: $H_{\l} = H_{0} + \l V$, where parameter $\l$ specifies the strength of the perturbation $V$ (which is considered to be a positive operator). The perturbed eigenstates $\ket{n}$ are now given by: $H_{\l} \ket{n} = E_{n} \ket{n}$, such that $ \lim_{\l \rightarrow 0}\ket{n} = \ket{n^{0}}$. From the standard treatment of stationary state perturbation theory, as discussed in Ref. \cite{sakurai1995modern}, one learns that the change in the spectrum due to perturbation is $\Delta_{n} = E_{n} - E_{n}^{0} = \avg{n^{0}}{\l V}{n}$, and the eigenstate $\ket{n}$ is expressible using the expansion: 
	\begin{align}\no
		\ket{n} = & \: \ket{n^0} + \l \sum_{k \neq n} \ket{k^0} \frac{V_{kn}}{E_{n}^{0} - E_{k}^{0}}\\ \non &+ \l^{2} \sum_{k \neq n} \sum_{l \neq n} \ket{k^0} \frac{V_{kl} V_{ln}}{(E_{n}^{0} - E_{k}^{0}) (E_{n}^{0} - E_{l}^{0})} \\& - \l^{2} \sum_{k \neq n} \ket{k^0} \frac{V_{kn} V_{nn}}{(E_{n}^{0} - E_{k}^{0})^{2}} + \mathcal{O}(\l^3),	
	\end{align}
	where $V_{mn} = \avg{m^0}{V}{n^0}$. From here it immediately follows that the change in the energy
	level can be expressed as an expansion:
	\begin{align} \no
		\Delta_{n} = & \: \l V_{nn} + \l^{2} \sum_{k \neq n} \frac{V_{nk} V_{kn}}{(E_{n}^{0} - E_{k}^{0})} \\ \non
		&+ \l^{3} \sum_{k \neq n} \sum_{l \neq n} \frac{V_{nk} V_{kl} V_{ln}}{(E_{n}^{0} - E_{k}^{0}) (E_{n}^{0} - E_{l}^{0})} \\ \non &- \l^{3} \sum_{k \neq n} \frac{V_{nk} V_{kn} V_{nn}}{(E_{n}^{0} - E_{k}^{0})^{2}} + \mathcal{O}(\l^4).
	\end{align}
	Interestingly the third term in the above expression consists of the contribution owing to the geometric phase $\gg_{V} = \text{Arg} \: V_{nk} V_{kl} V_{ln} = \text{Arg} (\avg{n^0}{V}{k^0} \avg{k^0}{V}{l^0} \avg{l^0}{V}{n^0})$.

	\subsection*{Manifestation in Scattering Theory}
	
	The generalised geometric phase $\gg_{V}$ due to the interaction potential also finds its manifestation in the scattering theory \cite{sakurai1995modern,gottfried2013quantum,cohen2019quantum}. To appreciate this let us consider the scattering problem wherein initially the particle is in the state $\ket{i} \equiv \ket{\vec{k}_{i}}$, which is an eigenstate of free Hamiltonian $H_{0} = \frac{\hat{p}^{2}}{2m}$. Eventually it interacts with a local potential $V(\vec{r})$ (which is a positive operator) and gets scattered to yield the final state $\ket{\j^{+}}$ after a large span of time. The time dependent perturbation theory in the interaction picture provides one with the scattering matrix, which is the amplitude for the particle to be found in $H_{0}$ eigenstate $\ket{n} \equiv \ket{\vec{k}_{n}}$ after time evolution for large time from the initial state $\ket{i}$. It is defined as: $S_{ni} = \avg{n}{U_{I}(\infty,-\infty)}{i} = \delta_{ni} - 2 \pi i \delta(E_{n} - E_{i}) T_{ni}$, where $E_{i(n)}$ is the initial(final) energy of the system. As defined earlier $U_{I}(t_f,t_i)$ is unitary time evolution operator defined in the interaction picture, with the identification $H_{\text{int}} = V$. The non-trivial $T$ matrix needed to understand the scattering matrix is defined via relation: $T_{ni} = \avg{n}{V}{\j^{+}}$.
	The $T$ matrix elements determine the scattering amplitude $f(\vec{k}_{n},\vec{k}_{i}) = - 4 \pi^{2} m \avg{n}{V}{\j^{+}}$, from which all the interesting consequences of the scattering process can be understood \cite{cohen2019quantum,gottfried2013quantum}. 
	The state $|\j^{+}\ran$ is iteratively found using the Lippmann-Schwinger equation \cite{sakurai1995modern,gottfried2013quantum}:
	\begin{align}
		\ket{\j^{+}} = \ket{i} + \frac{1}{E_{i} - H_{0} + i \epsilon} V \ket{\j^{+}}. 
	\end{align}
	The iterative substitution yields the Born series for $\ket{\j^+}$, often truncated to an appropriate order providing a Born approximation for $\ket{\j^+}$. Truncating the Born series at second order yields the expression for $|\j^{+}\rangle$ as:
	\begin{widetext}
		\begin{align}
			\ket{\j^+} = \ket{i} + \frac{1}{E_{i} - H_{0} + i \epsilon} V \ket{i} + \frac{1}{E_{i} - H_{0} + i \epsilon} V \frac{1}{E_{i} - H_{0} + i \epsilon} V \ket{i}.
		\end{align}
	\end{widetext}
	Let us define the state $\ket{\Delta i} = \frac{1}{E_{i} - H_{0} + i \epsilon} V \ket{i} + \frac{1}{E_{i} - H_{0} + i \epsilon} V \frac{1}{E_{i} - H_{0} + i \epsilon} V \ket{i}$ which essentially captures the contribution due to the scattering in $\ket{\j^+}$. This shows that the state $\ket{\j^{+}}$ is a linear superposition of two states $\ket{i}$ and $\ket{\Delta i}$, and one can ask about a possible interference phenomena due to this superposition. From the discussions in the earlier sections, one realises that the amplitude $\avg{i}{V}{\Delta i}$ can be used to study the interference effect. This amplitude can also be written as $\avg{i}{V}{\Delta i} = \avg{i}{V}{\j^{+}} - \avg{i}{V}{i}$ = $\frac{-1}{4 \pi^2 m} f(\vec{k}_{i}, \vec{k}_{i}) - \avg{\vec{k}_i}{V}{\vec{k}_i}$. Here the occurrence of the forward scattering amplitude $f(\vec{k}_{i}, \vec{k}_{i})$ is reassuring, since it has been long known that it captures the interference effect between the incoming wave and the outgoing wave in the scattering process. The optical theorem, which states that the total scattering cross section $\sigma_T$ is given by: $\text{Im} \: f(\vec{k}_{i}, \vec{k}_{i}) = \frac{|\vec{k}_{i}|}{4 \pi} \sigma_T$, is a testimony to this phenomenon. The forward scattering amplitude can be calculated using the above Born approximation to yield:
	\begin{align} \non
		\frac{-1}{4 \pi^2 m} f(\vec{k}_{i}, \vec{k}_{i}) &=  \avg{i}{V}{i} + \avg{i}{V \frac{1}{E_{i} - H_{0} + i \epsilon} V}{i} \\&
		+ \avg{i}{V \frac{1}{E_{i} - H_{0} + i \epsilon} V \frac{1}{E_{i} - H_{0} + i \epsilon} V}{i}.
	\end{align} 
	The second term in the above expression can be rewritten using the momentum eigenstates to read:
	\begin{align}
		\sum_{\vec{p},\vec{q}} \frac{\avg{\vec{k}_i}{V}{\vec{p}} \avg{\vec{p}}{V}{\vec{q}} \avg{\vec{q}}{V}{\vec{k}_{i}}}{(E_{i} - \frac{{\vec{p}}^2}{2m} + i \epsilon) (E_{i} - \frac{{\vec{q}}^2}{2m} + i \epsilon)},
	\end{align}
	clearly showing that it indeed receives a non-trivial contribution from the generalised geometric phase $\gg_{V} = \text{Arg} \: {\avg{\vec{k}_i}{V}{\vec{p}} \avg{\vec{p}}{V}{\vec{q}} \avg{\vec{q}}{V}{\vec{k}_{i}}}$. Thus one sees that within the purview of the second order Born approximation, both the forward scattering amplitude and the total scattering cross section contain a non-trivial contribution from the generalised geometric phase $\gg_{V}$.   
	
	\section{Summary}
	
	In this article, an operator generalisation of the concept of geometric phase for pure states is presented and its physical manifestations are discussed. The generalisation essentially stems from the fact that the interference phenomenon manifests in the average of observables, which paves the way to identify the argument of the matrix elements of an observable as a generalised relative phase. Interestingly, the difference of the generalised relative phase and the relative phase is found to be the argument of the weak value of the observable. From the generalised relative phase, following the trail of Pancharatnam, one is naturally lead to a generalised notion of geometric phase, which is defined at the least for a set of three states and a given observable. The geometric phase so defined is also found to hold in general for a collection of $N$ states, as also for the set of states that constitute a continuous curve in the Hilbert space. The generalised geometric phase in the latter case goes over to the usual geometric phase, as pioneered by Berry and others, when the observable is taken as identity operator. Remarkably it is observed that the generalised geometric phase is well defined in scenarios wherein the usual geometric phase is ill-defined. The notion of natural connection is also found to have an appropriate generalisation. 
	
	The notion of null phase curve is defined for the generalised geometric phase, and it is found that any two states in the Hilbert space of the system are always connected by a null phase curve. 
	We further find that the generalised relative phase, between any two states is expressible as a line integral of the generalised connection over the null phase curve connecting the two states. These null phase curves play a crucial role in ensuring that the generalised geometric phase is always expressible as a closed loop integral of the generalised connection, clearly showing that it is the (an)holonomy of the generalised connection.  
	
	It is well known that the usual geometric phase in the case of cyclic evolution manifests as a global phase acquired by the quantum state at the end of the evolution cycle. Here we find that the proposed geometric phase can also manifest in a similar fashion in a general quantum system, in particular, in scenarios wherein the usual geometric phase is not defined. It is found that the generalised geometric phase has several other manifestations in various problems of quantum mechanics. In the framework of time independent perturbation theory, it is observed that the generalised geometric phase contributes to the shift in the energy level due to the external perturbation. Interestingly in the scattering theory, the proposed geometric phase is found to non-trivially contribute to the forward scattering amplitude and to the total scattering cross section. 
	
	With this work, one sees that the occurrence and the notion of the geometric phase is much broader than it is currently understood, and was envisaged by its founders. It is well known that the geometric phase owes its existence to the symplectic structure of quantum ray space  \cite{vyas2023}, and so it is hoped that this generalisation would help shed some more light on the relationship of classical mechanics and quantum mechanics.  
	
	\appendix
	\section{Fibre bundle structure of the geometric phase}
	
	{Given the Hilbert space $\mathscr{H}$ for any physical system at hand, let us consider a subset $\mathscr{B} \subset \mathscr{H}$ such that $\mathscr{B} = \{ \ket{\j} \in \mathscr{H} | \amp{\psi}{\psi} = 1 \}$. Thus any given non-zero vector in $\mathscr{H}$ can be brought to a point in $\mathscr{B}$ by normalisation. While two unit vectors $\ket{\f}$ and $e^{i \l}\ket{\f}$ (for any real $\l$) represent mathematically distinct points on $\mathscr{B}$, they are physically indistinguishable. Thus it is physically meaningful to construct a space $\mathscr{R}$, called the \emph{ray space} or \emph{projective Hilbert space}, wherein $\ket{\f}$ and $e^{i \l}\ket{\f}$ are identified. So $\mathscr{R}$ is a quotient of $\mathscr{B}$ with respect to the equivalence relation $\ket{\f} \sim e^{i \l}\ket{\f}$, that is $\mathscr{R} = \mathscr{B}/\sim$ \cite{page1987,mukunda1993}. The elements of $\mathscr{R}$, called \emph{rays}, evidently can be represented by pure state matrices $\ket{\f}\bra{\f}$, giving rise to a projection map $\pi : \mathscr{B} \rightarrow \mathscr{R}$, defined as: $\pi (\ket{\f}) = \ket{\f}\bra{\f}$. Mathematically the triple $(\mathscr{B},\mathscr{R},\pi)$ constitutes a principal fibre bundle over the total space $\mathscr{B}$, base space $\mathscr{R}$ and with $U(1)$ as the structure group \cite{nakahara2003,page1987}. A fibre corresponding to a ray, consists of all the unit vectors projecting to the same ray.        
		
		Now suppose one is given a closed smooth curve $\mathcal{S}$ in $\mathscr{R}$, specified by $\r(l)$ (where $0 \leq l \leq L$) such that $\r(0)=\r(L)$. Evidently one can construct several different curves in $\mathscr{B}$ that all project to $\mathcal{S}$. Suppose one starts with some vector $\ket{\j(0)}$ on the fibre corresponding to $\r(0)$. One can then choose $\ket{\j(\epsilon)}$ on the fibre corresponding to $\r(\epsilon)$ (where $\epsilon$ is infinitesimally small), such that $\ket{\j(0)}$ and $\ket{\j(\epsilon)}$ are ``in-phase" to each other in the Pancharatnam sense, so that $\amp{\j(0)}{\j(\epsilon)}=1$. From here it immediately follows that $\ket{\j(0)}$ has to be chosen such that $\mathrm{Im} \avg{\j(0)}{\frac{d}{dl}}{\j(0)} = 0$. One can thus continue this procedure for all the other values of $l$, and get a unique curve $\ket{\j(l)}$ for which the natural connection vanishes: $A_{\j}(l) = \mathrm{Im} \avg{\j(l)}{\frac{d}{dl}}{\j(l)} = 0$. Such a curve is called the horizontal lift of $\mathcal{S}$\cite{mukunda1993}. This shows that the origin of the connection $A_{\j}(l) = \mathrm{Im} \avg{\j(l)}{\frac{d}{dl}}{\j(l)}$ essentially stems from the idea of comparing the two nearby vectors in $\mathscr{B}$ employing the inner product, justifying the adjective `natural' to its name.        
		
		Interestingly the horizontal lift $\ket{\j(l)}$ that one obtains following the above procedure need not necessarily be a closed curve $\ket{\j(L)} \neq \ket{\j(0)}$, but can admit an accumulated phase $\gamma$: $\ket{\j(L)} = e^{-i \gamma}\ket{\j(0)}$. From (\ref{gpcont}) one can immediately see that $\gamma$ is nothing but the geometric phase corresponding to the curve $\mathcal{S}$. It must be noted that in general if one demands that the curve $\ket{\j(l)}$ be a closed curve so that $\ket{\j(L)} = \ket{\j(0)}$, then one can not ensure that such a curve also be a horizontal lift.   
		
		\section{Fibre bundle structure behind the generalised geometric phase}
		
		Consider that one is given some physical system at hand and its corresponding \emph{N} dimensional Hilbert space $\mathscr{H}$. In such a system let us also consider some observable $\oo$ of interest, which admits distinct real eigenvalues $\lambda_j$: $\oo \ket{e_j} = \lambda_j \ket{e_j}$, where the eigenvectors $\ket{e_j}$ form a complete orthonormal set $\amp{e_i}{e_j} = \delta_{ij}$. For simplicity we shall also assume that none of the eigenvalues $\lambda_j$ are zero. 
		
		For any given non-zero state $\ket{\j}$, the average or the expectation value of $\oo$ is given by $\oo_{\j} = \frac{\avg{\j}{\oo}{\j}}{\amp{\j}{\j}}$. Let $\mathcal{K}_{\oo}$ be the set of states for which this average vanishes: $\mathcal{K}_{\oo} = \{ \f \in \mathscr{H} | \avg{\f}{\oo}{\f} = 0  \}$. Resolving the state $\ket{\f}$ into the basis $\ket{e_j}$: $\ket{\f} = \sum_{j} z_{j} \ket{e_j}$, immediately yields: $\sum_{j} \lambda_j |z_j|^2 = 0$. This shows that $\mathcal{K}_{\oo}$ is a homogeneous quadric surface, a higher dimensional generalisation of conical surface. 
		
		Inorder to define the generalised geometric phase, we shall only work with vectors that all lie in the set $\mathscr{H} - \mathcal{K}_{\oo}$. Instead of working with the usual normalisation $\amp{\j}{\j} = 1$, let us work with the normalisation: $\amp{\j}{\j} = |\oo_{\j}|^{-1}$. With this normalisation any vector in $\mathscr{H} - \mathcal{K}_{\oo}$ can be brought to either of the sets $\mathscr{B}^{\pm}_{\oo} = \{ \f \in \mathscr{H} - \mathcal{K}_{\oo} | \: \: \avg{\f}{\oo}{\f} = \pm 1 \}$. Evidently the two sets $\mathscr{B}^{\pm}_{\oo}$ are disjoint. Any state $\ket{\f}$ can be resolved into the basis $\ket{e_j}$: $\ket{\f} = \sum_{j} z_{j} \ket{e_j}$, so that one has: $\sum_{j} \lambda_j |z_j|^2 = \pm 1$. This shows that the sets $\mathscr{B}^{\pm}_{\oo}$ are smooth quadric surfaces, which are higher dimensional generalisations of conic sections. If all the eigenvalues are either positive or negative, one gets surfaces which are generalised ellipsoids. Else one has the surfaces as generalised hyperboloids.     
		
		This sets the ground for defining the ray spaces $\mathscr{R}^{\pm}_{\oo}$ respectively corresponding to the sets $\mathscr{B}^{\pm}_{\oo}$ as the quotient $\mathscr{R}^{\pm}_{\oo} = \mathscr{B}^{\pm}_{\oo}/\sim$, where the equivalence relation $\ket{\f} \sim e^{i \theta} \ket{\f}$ (for any real $\theta$) is defined for any $\ket{\f}$ in $\mathscr{B}^{\pm}_{\oo}$. One thus has a pair of  projection maps $\pi^{\pm}: \mathscr{B}^{\pm}_{\oo} \rightarrow \mathscr{R}^{\pm}_{\oo}$ respectively as: $\pi^{\pm}(\ket{\f}) = \r_{\f} = \ket{\f}\bra{\f}$. It is evident that the pair of triples $(\mathscr{B}^{+}_{\oo}, \mathscr{R}^{+}_{\oo},\pi^{+})$ and  $(\mathscr{B}^{-}_{\oo}, \mathscr{R}^{-}_{\oo},\pi^{-})$ form separate principle fibre bundles respectively with the total space as $\mathscr{B}^{\pm}_{\oo}$, base space as $\mathscr{R}^{\pm}_{\oo}$, and $U(1)$ as the structure group. It must be emphasised that these two structures in general are fundamentally different from the the triple $(\mathscr{B},\mathscr{R},\pi)$ discussed earlier. 
		
		For example, in the case of two level quantum system, the space $\mathscr{B}$ is formed by the vectors $\ket{\j} = e^{i\chi} (\cos \theta \ket{0} + e^{i \f} \sin \theta \ket{1})$, for real values of $\chi$, $\theta$ and $\f$. Ignoring the global phase $e^{i\chi}$, one immediately sees that each $\ket{\j}$ corresponds to a point on the Bloch sphere, which is parameterised by two angles $\theta$ and $\f$, which forms the ray space $\mathscr{R}$. However if now one considers $\oo = \sigma_{z}$, then the states that belong to $\mathscr{B}_{\sigma_{z}}^{+}$ are $\ket{\J} = e^{i\chi} (\pm \cosh \theta \ket{0} + e^{i \f} \sinh \theta \ket{1})$. Ignoring the global phase $e^{i\chi}$ one sees that each vector $\ket{\J}$ now corresponds to a point on a two-sheeted hyperboloid, which is defined in $\mathrm{R}^{3}$ by the coordinates $(\pm \cosh \theta, \sinh \theta \sin \f, \sinh \theta \cos \f )$. This is the ray space $\mathscr{R}^{+}_{\sigma_{z}}$. It can be checked that proceeding similarly, the ray space $\mathscr{R}^{-}_{\sigma_{z}}$ turns out to be a single-sheeted hyperboloid formed out of rays corresponding to the vectors $\ket{\J} = e^{i\chi} ( \sinh \theta \ket{0} + e^{i \f} \cosh \theta \ket{1})$. It is easy to see that the non-compact nature of the ray space arises due to the positive and negative eigenvalues of $\sigma_{z}$. 
		
		Now one can consider a closed smooth curve in $\mathscr{R}^{+}_{\oo}$ say $S$ specified by $\r_\j(l) = \ket{\j(l)} \bra{\j(l)}$, ($0 \leq l \leq L$) such that $\r_\j(0) = \r_\j(L)$. Corresponding to such a curve one can ask if it is possible to construct a horizontal lift of it in $\mathscr{B}^{+}_{\oo}$. In the usual geometric phase case, in order to compare two infinitesimally separated vectors $\ket{\j(l)}$ and $\ket{\j(l+\epsilon)}$ on a smooth curve, the inner product $\amp{\j(l)}{\j(l+\epsilon)}$ was employed, and the ``in-phase" condition of transport dictated that the natural connection $\mathrm{Im} \avg{\j(l)}{\frac{d}{dl}}{\j(l)}$ should vanish. In this case, let us generalise the earlier definition, and compare $\ket{\j(l)}$ and $\ket{\j(l+\epsilon)}$ belonging to a smooth curve in $\mathscr{B}^{+}_{\oo}$, with the ``$\oo$ in-phase'' criteria.  Demanding that $\ket{\j(l)}$ and $\ket{\j(l+\epsilon)}$ should be ``$\oo$ in-phase'' with one another, immediately implies that $\avg{\j(l)}{\oo}{\j(l+\epsilon)}$ must be real and positive. This in turn dictates that the generalised connection ${A}^{\oo}(l) = \mathrm{Im} \avg{\j(l)}{\oo \frac{d}{dl}}{\j(l)}$ vanishes. One can now construct a horizontal curve ensuring that the generalised connection vanishes. From (\ref{ogpcont}) it is immediately clear that for such a curve the total accumulated phase $\ket{\j(L)} = e^{-i \gamma} \ket{\j(0)}$ is the generalised geometric phase $\gg_{\oo}(0,L)$.           
		
		To further understand the geometry behind the generalised geometric phase, let us consider a closed smooth curve $C$ in $\mathscr{B}^{+}_{\oo}$, given by $\ket{\f(l)}$ ($0 \leq l \leq L$). Without loss of generality, we can express it in the eigenbasis of $\oo$ as: $\ket{\f(l)} = \sum_{j} z_{j}(l) \ket{e_j}$, which in turn gives us the generalised geometric phase as:
		\begin{align*}
			\gg_{\oo}(0,L) = \oint_{C} dl {A}^{\oo}(l) = \oint_{C} \sum_{j} \mathrm{Im} ( \l_{j} z^{\ast}_{j} d z_{j}).
		\end{align*}
		Employing the Stokes theorem this can be written as an integral over some surface $S$ bounded by $C$: $\gg_{\oo}(0,L) = \int_{S} d {A}^{\oo}$, where the two-form $d {A}^{\oo}$ is:
		\begin{align*}
			d{A}^{\oo} = \sum_{j} \l_{j} (dp_{j} \wedge dx_{j}),
		\end{align*}
		where $z_{j} = \frac{1}{\sqrt{2}}(x_j - i p_j)$. For the usual geometric phase, $\l_{j}$s are all unity, so that $\gg(0,L)$ can be understood as a sum of area elements $dp_{j} \wedge dx_{j}$. The generalised geometric phase, on the other hand, is a weighted sum of the area elements $dp_{j} \wedge dx_{j}$ with $\l_{j}$s as the weights.}

	\begin{acknowledgments}
		The author is indebted to Profs. N. Mukunda, J. Samuel and D. Roy for several beneficial and enlightening discussions on various aspects of the geometric phase. The author also thanks Profs. M. V. Berry, R. Nityananda and P. K. Panigrahi for encouraging conversations.
	\end{acknowledgments}
	
	
	%

\end{document}